\documentclass[doublecol]{epl2} 
\usepackage{amssymb}
\newcommand{\bea}{\begin{eqnarray}}
\newcommand{\eea}{\end{eqnarray}}
\newcommand{\beq}{\begin{equation}}
\newcommand{\eeq}{\end{equation}}

\title{Phase-locking transition of coupled low-dimensional superfluids}
\shorttitle{Phase-locking transition of coupled low-dimensional superfluids} %Insert here a short version of the title if it exceeds 70 characters

\author{L.~Mathey\inst{1} \and  A.~Polkovnikov\inst{2} \and A.~H.~Castro~Neto\inst{2}}
\shortauthor{L.~Mathey,  A.~Polkovnikov, \and A.~H.~Castro~Neto}

\institute{                    
  \inst{1} Physics Department, Harvard University, Cambridge, MA 02138\\
  \inst{2} Department of Physics, Boston University, 590 Commonwealth Ave., Boston, MA 02215
}

\abstract{
We study the phase-locking transition of two coupled low-dimensional
superfluids, either two-dimensional superfluids at finite
temperature, or one-dimensional superfluids at zero temperature. We
find that the superfluids have a strong tendency to phase-lock. The
phase-locking is accompanied by a sizeable increase of the
transition temperature $T_{c}$ (in 2D systems) of the resulting
double-layer superfluid to thermal Bose gas transition, compared
 to the Kosterlitz-Thouless temperature $T_{KT}$ of 
 the uncoupled 2D systems, 
 which suggests a plausible way of observing
the Kibble-Zurek mechanism in two-dimensional cold atom systems by
rapidly varying the tunneling rate between the superfluids. If there
is also interaction between atoms in different layers present
 we find
additional phases, while no sliding phase,
  characterized by order or quasi long range order (QLRO) either in
 the symmetric or the antisymmetric sector of the system.
}

\begin{document}

\maketitle

\section{Introduction}
Since the realization of the Mott insulator transition
\cite{greiner}, remarkable progress has been made in controlling and
manipulating ensembles of ultra-cold
atoms~\cite{stoeferle}, which was followed by a number
of experiments to create and study more and more sophisticated
many-body effects, such as fermionic superfluids \cite{BECBCS},
one-dimensional strongly correlated Fermi and Bose
systems~\cite{Tonks}, or noise correlations in
interacting atomic systems~\cite{ehud, noise}. 
%These developments
%established the notion of 'engineering' many-body states in a
%tunable environment, i.e. manipulating ensembles of ultra-cold atoms
%in optical lattices. 
 An intriguing new direction in studying
low-dimensional strongly correlated systems was taken with the
realization and observation~\cite{zoran} of the Kosterlitz-Thouless
(KT) transition~\cite{chaikin-lubensky}. In this experiment the
interference amplitude between
 two
 %a pair of 
 independent two-dimensional
(2D) Bose systems was studied as a function of temperature. This
analysis revealed the jump in the superfluid stiffness (see also
Ref.~\cite{pad}) and the emergence of unpaired isolated
vortices as they crossed the phase transition.

In another interesting recent experiment Sadler {\em et. al.}
observed spontaneous generation of topological defects in the spinor
condensate after a sudden quench (i.e. a rapid, non-adiabatic
 ramp) through a quantum phase
transition~\cite{kurn}. A similar experiment in a double-layer
system was reported in Ref.~\cite{scherer}. The topological
defects are generated~\cite{kibble} at a density which is related to
the rate at which the transition is crossed~\cite{zurek}. Later it
was argued that the dependence of the number of such defects on the
swipe rate across a quantum critical point can be used as a probe of
the critical exponents characterizing the phase
transition~\cite{kbz}. 
 This Kibble-Zurek (KZ) mechanism 
 was originally considered as an early universe scenario 
%to have occurred just after the Big Bang, when the emerging phase of
%the vacuum became disordered, giving rise to a large number of
%topological defects such as 
 creating cosmic strings, which 
 would serve as an 
 %the primal
 ingredient for the formation of galaxies \cite{kibbleremark}. 
 Cold atom systems appear
to be a very  
 %the most
 suitable laboratory for performing such
``cosmological experiments'', since these systems are
highly tunable and well isolated from the environment. So far the
experiments and the theoretical proposals addressed the KZ 
scenario across a quantum phase transition. The main reason is that
it is generally hard to cool such systems sufficiently fast to
observe non-equilibrium effects. In this work 
 we provide 
 an example of a
particular system 
 where 
this difficulty can be easily
overcome by quenching the transition temperature $T_c$ instead of
$T$. Thus the relevant ratio $T/T_c$ can be tuned with an arbitrary
rate and the KZ mechanism can be observed. 
%Specifically we
%examine a problem of two coupled superfluids (SF). Our proposal for
%the realization of the KZ effect in coupled SFs is as
%follows: one prepares two or several decoupled systems slightly
%above the KT transition so that they are in the disordered phase,
%then one switches on the tunneling between them at some finite rate.
%As we show below this tunneling rapidly increases the KT transition
%temperature $T_{KT}$ and the system attempts to create long-range
%order (LRO).
 Specifically, we
examine a system of two superfluids (SF):
%  Our proposal for
%the realization of the KZ effect in coupled SFs is as
%follows: one prepares two or several decoupled systems slightly
%above the KT transition so that they are in the disordered phase,
%then one switches on the tunneling between them at some finite rate.
As we show below, by turning on tunneling 
 between the two systems
 the transition
temperature
  increases rapidly, 
 and the system attempts to create long-range
order (LRO).
 However, in this process, defects in the SF
phase are created, which develop into long-lived vortex-anti-vortex
pairs or in finite system unbalanced population between vortices and
anti-vortices. We note that because the systems are isolated and
there is no external heat bath, the temperature itself also changes
due to the quench. However, the long-wavelength fluctuations
relevant for the KT transition are only a small subset of all
degrees of freedom, majority of which are only weakly affected by
small inter-layer tunneling. So we believe that the change of the
$T_{c}$ is the main effect of the quench.

\begin{figure}
\onefigure[width=8.6cm]{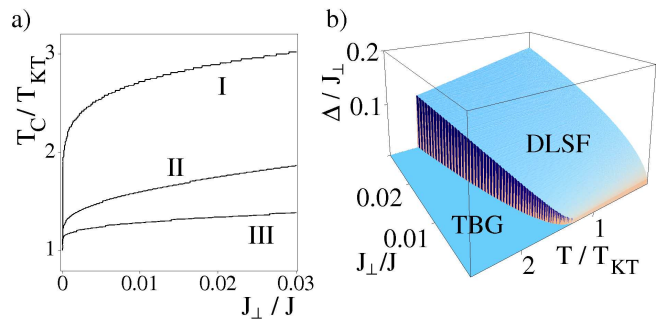}
\caption{\label{PD} (a) Critical temperature $T_c$ of the 
 DLSF-TBG transition (in units of $T_{KT}$)
for different values of $A_1/J$: $10^{-3}, 0.1, 0.4$ (I--III), and
 for $J_{\rm int} =0$. (b)
Energy gap in the anti-symmetric sector (in units of $J_{\perp}$) as
a function of $J_{\perp}/J$ and temperature (in units of $T_{KT}$).
We have set $A_1/J=0.1$
 and $J_{int}=0$.}
\end{figure}

In this work we consider two SFs coupled via tunneling and/or
interactions. In the experiments the hopping or tunneling rate
between two systems can be tuned to a high precision
\cite{Tonks,zoran,joerg,zoran1}. Interactions between the atoms in
different systems can either be realized in ensembles of polar
molecules or by using mixtures of two hyperfine states, where the
tunneling rate is controlled by an infrared light
source~\cite{phase}, which induces 
 spin-flipping between the hyperfine states. 
 In this case the atoms in different states
naturally interact with each other since they are not physically
separated in space. 
%Our analysis applies to fermionic SFs as well,
%where the hopping term would correspond to the hopping of pairs of
%atoms. 
 We also discuss the analogous one-dimensional (1D), zero
temperature lattice systems. The main results of our analysis are
the phase diagrams of coupled SFs in Fig.~\ref{pdint} (2D) and
 Fig. \ref{pdint1} (1D), the behavior of $T_{c}$ and the energy gap
shown in Fig.~\ref{PD}, as well as
 the proposal of realizing the KZ mechanism by
 switching on the tunneling between two SFs.
\begin{figure}
\onefigure[width=5.3cm]{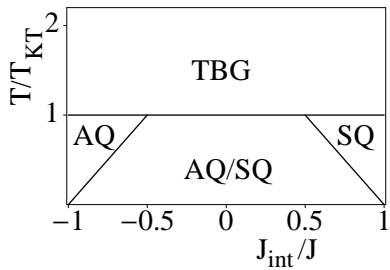}
\caption{\label{PD_Jint} Phase diagrams of two 2D SFs, coupled
 through a term of the form ${\mathcal S}_{12}$, Eq. (\ref{S12}), 
in terms of $J_{int}/J$ and $T/T_{KT}$. 
 For low temperatures we find antisymmetric quasi-order (AQ) and/or 
 symmetric quasi-order (SQ), which either simultaneously undergo a KT
 transition due to single vortices  (AQ/SQ to thermal Bose gas (TBG) phase), 
 or individually due to correlated vortex
 pairs: symmetric (anti-symmetric) vortex pairs drive the AQ/SQ to AQ (SQ) transition.
 %, symmetric
% vortex pairs drive the AQ/SQ to AQ transition.
}
\end{figure}
\section{2D superfluids}
%Let us start the analysis considering 
 In this section we consider 
two 2D SFs, each characterized
by a KT temperature $T_{KT}$. 
 %It is convenient to 
 We write the bosonic
operators $b_{1/2}$ in the two layers in a phase-density
representation~\cite{chaikin-lubensky,gnt}, $b_{1/2}\sim
\sqrt{\rho_{1/2}} \exp(i \phi_{1/2})$, where $\rho_{1/2}$ are 
 the density operators of the two 
 systems, and 
 %the
 $\phi_{1/2}$ the phases. 
 The low-momentum fluctuations
of the phase fields are described by Gaussian contributions to the
Hamiltonian $\mathcal H_0$. Because of the formal analogy between
the quantum 1D and thermal 2D systems~\cite{giamarchi_book} we adopt
the quantum terminology throughout the paper and refer to the ratio
of the Hamiltonian and the temperature as the action. Then
\beq
{\mathcal S}_{0}\equiv \frac{{\mathcal H}_0}{T} = \frac{J}{2 T} \int d^2r
[(\nabla \phi_1)^2+(\nabla \phi_2)^2] \,.
\label{S0}
\eeq
The energy scale $J$ here is related to $T_{KT}$ by $J=2
T_{KT}/\pi$.
Besides these long-wavelength fluctuations, the system also contains
additional degrees of freedom, vortex-anti-vortex
pairs~\cite{chaikin-lubensky}. The corresponding term in the action
  is expressed through the dual fields $\theta_{1,2}$~\cite{gnt}:
\bea\label{S1}
{\mathcal S}_1 & = &  \frac{2 A_1}{T} \int \frac{d^2 r}{(2 \pi\alpha)^2}
[\cos(2\theta_1) + \cos(2\theta_2)] \, ,
\eea
where $\alpha$ is a short-distance cut-off of the size of the vortex
core, and 
 %the coefficient
 $A_1$ 
 %of this term 
 is proportional to the
 single-vortex fugacity: $A_1\sim J\exp(-J/T)$,  
 % Here, and throughout the paper, 
 where we assume 
 %the two 
 both SFs 
 %are 
%each
 % characterized by 
 to have the same effective parameters $J$ and
 $A_1$. 
Operators of the type
$\exp(2 i \theta)$ create kinks in the 
 %phase 
field $\phi$: $\exp(-2
i \theta(x))\phi(x')\exp(2 i \theta(x)) \sim \phi(x') + 2\pi \Theta(x
-x')$, $\Theta(x)$ being the step function, which 
 %is equivalent 
 corresponds to
the effect of vortices in the original 2D problem 
 %(for more details
 (Ref.~\cite{giamarchi_book}, p. 92).

In addition the two systems are coupled by a hopping term $\sim
t_\perp b_1^\dagger b_2 + h.c.$, which results in the following
contribution to the action:
\bea\label{Sperp}
{\mathcal S}_\perp & = &  \frac{2 J_\perp}{T} \int \frac{d^2 r}{(2 \pi\alpha)^2} \cos(\phi_1-\phi_2),
\eea
where the bare value of $J_\perp$ corresponds approximately to
$t_\perp \rho_0$. In principle, the hopping term is modified by the
vortex contributions, however, these corrections are always
irrelevant under renormalization group (RG). 

For most of the discussion in this paper
 we use the symmetric and anti-symmetric combinations of 
%the
% fields 
 $\phi_{1/2}$ and $\theta_{1/2}$:
\beq
\phi_{s/a} = (\phi_1 \pm \phi_2)/\sqrt{2} ,\quad
\theta_{s/a} = (\theta_1 \pm \theta_2)/\sqrt{2} \, .
\label{spincharge}
\eeq
 Written in 
 %terms of
 these fields, 
%the quadratic part
% of the action introduced sofar, i.e. 
the term ${\mathcal S}_0$ in
 Eq. (\ref{S0}) is again a sum of Gaussian models, now in 
 the fields $\phi_s$ and $\phi_a$,
 % instead of $\phi_1$ and $\phi_2$,
 with the same energy scale $J$.
 However, we will consider a broader class of actions,
 % in this paper,
 in which the energy scales of the symmetric and anti-symmetric sector
  differ. We include the following term in the action:
\bea\label{S12}
{\mathcal S}_{12} & = &  \frac{J_{{\rm int}}}{T} \int d^2r \nabla \phi_{1}\nabla\phi_2
\eea
 With this, the quadratic part of the action
 %, i.e.  Eqs.~(\ref{S0}) and
%(\ref{S12}), 
%written in terms of the fields $\phi_{s/a}$, 
 is given by:
\bea
{\mathcal S}_{0} + {\mathcal S}_{12} & = & \frac{J_s}{2 T} \int d^2r
 (\nabla \phi_s)^2 + \frac{J_a}{2 T} \int d^2r
(\nabla \phi_a)^2 \, ,
\label{Squadr}
\eea 
 where $J_s$ and $J_a$ are 
 given by $J_{s/a}=J \pm J_{{\rm int}}$.

 We now motivate the existence of such a term ${\mathcal S}_{12}$ in
 ultracold atom systems, by considering two BECs coupled
 by a short-range density-density interaction.
 %Following the usual treatment of weakly interacting 
 %bosons, 
 Starting from a Hamiltonian of the 
 form $H = \sum_k [\epsilon_{\mathbf k} b^\dagger_{\mathbf k} b_{\mathbf k} 
 + (g/2V) \rho_{\mathbf k}^\dagger \rho_{\mathbf k}$],
 where $b_{\mathbf k}$ is the boson operator, $\epsilon_{\mathbf k}$ the free 
 dispersion $\epsilon_{\mathbf k}={\mathbf k}^2/2m$, $g$ is the interaction
  strength of the contact interaction, $V$ the volume,
% of
% the system, 
 and $\rho_{\mathbf k}$ is the density operator of momentum 
 ${\mathbf k}$,
 given by $\rho_{\mathbf k}=\sum_{\mathbf p} b^\dagger_{\mathbf p} b_{\mathbf p+\mathbf k}$,
 we assume that the zero momentum  mode is macroscopically
occupied, and 
 formally replace the 
 operator $b_0$ by a number, 
 $b_0 \rightarrow \sqrt{N_0}$, where
 $N_0$ is the number of condensed atoms which is 
 comparable to the total atom number $N$, i.e. $N_0 \lesssim N$.
 Next we keep all terms that
 are quadratic in $b_\bold{k}$ (with $\bold{k}\neq0$), and
 perform a Bogoliubov transformation, given by:
 $b_\bold{k} = u_\bold{k} \beta_\bold{k} + v_\bold{k} \beta^\dagger_\bold{-k},$
 to diagonalize the Hamiltonian.
 The 
 eigenmodes $\beta_\bold{k}$ have a
 dispersion relation 
 %given by 
  $\omega_{\mathbf k} = \sqrt{\epsilon_{\mathbf k}(\epsilon_{\mathbf k} +2 g n)}$,
 with $n$ being the density $N/V$.
  The low-${\bf k}$ limit is given by
 $\omega_\bold{k}^2\sim v^2|{\bf k}|^2$, 
 with $v=\sqrt{g n/m}$, which 
 corresponds to the contribution in Eq. (\ref{S0}) of the action.
  Next, we consider 
 %two copies of this system, i.e. we have
 the sum of two copies of the previous Hamiltonian with boson operators
 $b_{1/2}$. 
 In addition we consider an interaction 
  $H_{12}= g_{12}/V \sum_{\mathbf k} \rho_{1, \mathbf k}^\dagger \rho_{2, \mathbf k}$,
 where the density operators 
 $\rho_{1/2, {\mathbf k}}$ are 
given by $\rho_{1/2, \mathbf k}=\sum_{\mathbf p} b^\dagger_{1/2, \mathbf p} b_{1/2, \mathbf p+\mathbf k}$.
 Following the same procedure as before, we find two eigenmode
 branches, corresponding to 
 in-phase and out-of-phase superpositions of the modes of
 each condensate, with
 the 
 dispersions $\omega_{s/a, \bold{k}}^2\sim v_{s/a}^2|{\bf k}|^2$,
 with the velocities  $v_{s/a}=\sqrt{(g \pm g_{12}) n/m}$.
 Therefore, for this example, the energy scale $J_{int}$ is related to
 $g_{12} n/m$, 
  which would be of similar order as $J$ for a system
 interacting via contact interaction, for small temperatures.
 This discussion only applies to the weakly interacting limit of a true 
condensate. However, it demonstrate that 
 %the presence of
 a density-density contact interaction term can lead to a substantial 
 energy splitting of the
 in-phase and out-of phase modes. 

\begin{figure}[t]
\onefigure[width=5.1cm]{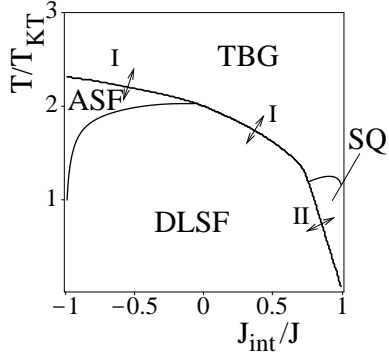}
\caption{\label{pdint} Phase diagram, temperature (in units of
$T_{KT}$) versus interaction (in units of $J$). We assume $J_\perp
/J\sim 10^{-3}$ and $A_1/J \sim 10^{-3}$. DLSF: double layer
superfluid; TBG: thermal Bose gas; ASF: anti-symmetric superfluid;
SQ: symmetric quasi-order. The order of the transition
lines are either first (I), second order (II), or KT (thin lines).}
\end{figure}

% If the two systems have
% density-density interactions then there is an additional
% contribution to the action:
% \bea\label{S12}
% {\mathcal S}_{12} & = &  \frac{J_{{\rm int}}}{T} \int d^2r \nabla \phi_{1}\nabla\phi_2
% \eea
% We note that even if not present at the onset this term will still
% be generated by the RG transformations  as long as there is nonzero
% tunneling between the two systems.

% It is convenient to change the variables to the symmetric and anti-symmetric combinations:
% \beq
% \phi_{s/a} = (\phi_1 \pm \phi_2)/\sqrt{2} ,\quad
% \theta_{s/a} = (\theta_1 \pm \theta_2)/\sqrt{2} \, .
% \label{spincharge}
% \eeq
% The quadratic part of the action given by Eqs.~(\ref{S0}) and
% (\ref{S12}) decouples into the sum of symmetric and anti-symmetric
% parts with the parameters $J_s$ and $J_a$, characterizing each of
% these sectors ($J_{s/a}=J \pm J_{{\rm int}}$).

Finally, in addition
%to the terms describing 
 to single 
 vortices in each SF,
 % of the systems separately,
we have to consider the possibility of correlated vortex pairs, 
 i.e. one vortex in each layer at the same location of either
 the same 
 %vorticity
 or of opposite vorticity.  
 We will refer to these vortex configurations as 
 {\it symmetric} or {\it anti-symmetric vortex pairs}, respectively. 
 These 
 %types of
 excitations appear as the following terms
 in the action:
%
% Even if not present at the onset, these terms are
%generated under the RG flow at second order:
%
%
\bea\label{Srs}
{\mathcal S}_{s,a} & = & \frac{2 A_{s,a}}{T} \int \frac{d^2 r}{(2
\pi\alpha)^2} \cos(2\sqrt{2}\,\theta_{s,a}).
\eea
These 
 correlated vortex terms, 
 %coupled vortices form 
 which describe new
degrees of freedom,
 %  corresponding to 
 can be the most relevant non-linear
terms in the action, 
 %, for some of the parameter regime under consideration.  
%
 %The importance and relevance of these terms follows 
 which derives from the
possibility that the vortices in different layers interact with each
other, through the
 terms (\ref{Sperp}) and (\ref{S12}). 
% Indeed, it is easy to see that 
  The effect of these terms is the following:
 At low temperatures the energy
between two single vortices of opposite
 vorticity due to tunneling 
 %term 
 grows as the
square of the distance D between them,
 i.e. as $J_\perp (D/\alpha)^2$. 
 As a result, the tunneling term attempts
 to confine vortices of opposite vorticity, leading
 to phase-locking between the layers, which we describe further
 later on.
%
 %Furthermore, 
 The interaction ${\mathcal S}_{12}$ changes
 the energy of correlated vortex pairs as follows: 
 %in the following way:
 The energy of a single vortex is given by
  $2\pi J\log L/\alpha$, where $L$ is the system size,
  whereas
  a symmetric/anti-symmetric vortex pair has
 an energy of 
 $4\pi(J\pm J_{int})\log L/\alpha$.
  Therefore, symmetric vortex pairs are the lowest energy vortex
 excitations for $J_{int}<-J/2$, whereas for $J_{int}>J/2$ anti-symmetric
 vortex pairs are the lowest energy excitations.
 % with lowest energy.
 As we will see below, in these regimes correlated vortex pairs 
 %excitations
 drive transitions to phases, in which one sector
 is (quasi-)SF whereas the other is disordered. 
 % In the RG approach shown below, 
 We  will also see that
 these terms are
generated under the RG flow, 
 % at second order, 
 even if not present at the onset.

 We note that a similar system has been studied in \cite{cazalilla}. Here,
 we consider a larger class of systems by including
 the interaction term (\ref{S12}), which in turn requires us to include
 the correlated vortex excitations (\ref{Srs}). These terms give rise
 to additional phases as we will see in the following.

%These coupled vortices form new
%degrees of freedom corresponding to the most relevant non-linear
%terms in the action.

Next we analyze our system within the RG approach. 
 This RG flow is perturbative in the vortex fugacities 
 $A_1$, $A_{s}$, and $A_a$, and the tunneling energy $J_\perp$,
 and therefore applies to the weak-coupling limit 
 (in particular 
 %to the limit
 $J_\perp\rightarrow +0$).
 At second order
%in the vortex fugacity and $J_\perp$  
 the flow equations are given
by \cite{giamarchi}:
\bea
\label{RG}
&&\frac{d J_\perp}{d l}=\left(2-\frac{T}{2\pi J_a}\right) J_\perp \, ,
\label{JperpEqn}\\
&&\frac{d A_s}{d l}= \left(2-2\pi \frac{J_s}{T}\right) A_s +
\alpha_3\frac{A_1^2 (J_a -J_s)}{2 T^2} \, ,
\\
&&\frac{d A_a}{d l}= \left(2-2\pi \frac{J_a}{T}\right) A_a +
\alpha_3\frac{A_1^2 (J_s -J_a)}{2 T^2} \, ,
\\
&&\frac{d A_1}{d l}=\left(2-\frac{ \pi(J_s+ J_a)}{2T} + \alpha_3\frac{A_s J_s
    + A_a J_a}{T^2}\right) A_1 \, ,
\label{A1Eqn}
\\
&&\frac{d J_a}{d l}=\alpha_2 \Big(\frac{J_\perp^2}{4\pi^4 J_a} -
4\frac{A_a^2}{ T^4}  J_a^3 - \frac{A_1^2}{2 T^4}
(J_s+J_a)J_a^2\Big), \phantom{XX}
\\
&&\frac{d J_s}{d l}= - \alpha_2 \Big(2\frac{A_s^2}{ T^4}  J_s^2 +
\frac{A_1^2}{4 T^4} (J_s+J_a)J_s\Big) 2 J_s \, .
\eea
The coefficients $\alpha_{2/3}$ are non-universal parameters that
appear in the RG procedure~\cite{SG}, and which do not  affect the
results qualitatively. For consistency, we have to expand the
right-hand site of the above equations up to second order, around
the resulting Gaussian fixed point: $J_{s/a} = J \pm J_{{\rm int}} +
j_{s/a}$. We emphasize again that $J_{int}$ near the fixed point can
be generated by RG and be nonzero even if it is not present at the
onset.

Before we consider the full RG flow, we consider the 
 simpler case of no tunneling, i.e. 
 we solve the RG equations while setting 
$J_\perp =0$.
 In Fig. \ref{PD_Jint} we show the phase diagram of two 2D SFs coupled
 by 
 %a term of the form 
 ${\mathcal S}_{12}$, Eq. (\ref{S12}).
  Such a system would be realized
 by a 2D mixture of bosonic atoms in two different hyperfine
 states, interacting via some short-range potential. 
 The order parameters we consider are $O_s(x)=b_1(x) b_2(x)$ 
and $O_a(x)=b_1^\dagger(x) b_2(x)$. 
 To obtain the phase diagram we consider
 the correlation functions of each of these order parameters,
 which can either scale algebraically or exponentially.
 In Fig. \ref{PD_Jint} we refer to algebraic scaling
 of $O_s(x)$ 
 %in the symmetric sector 
as symmetric quasi-order (SQ), 
 and 
 of $O_a(x)$
%in the anti-symmetric sector 
 as anti-symmetric quasi-order (AQ). 
 In each of the sectors a KT transition marks the 
 transition from the algebraic to the exponential regime,  
  which occur either simultaneously and are driven by single-vortex 
 excitations, or at different temperatures and are driven
 by correlated vortex pairs.
  As a result we find four regimes:
 At temperatures above $T_{KT}$, both sectors are disordered, giving
 rise to a thermal Bose gas (TBG) phase. For temperatures
 below $T_{KT}$, and for a wide range of $J_{int}$, we find
 that both sectors 
 are quasi SF
 %show quasi-superfluidity 
 (AQ/SQ), which
 is the only phase in which the correlation function of the
 single boson operators show algebraic scaling.
 We also find regimes in which only one sector shows
 algebraic scaling, whereas the other is disordered (AQ and SQ).
  %In both of these regimes the single boson operators themselves
  %show exponential scaling. 
  From the perspective of vortices, the
 TBG phase is a gas of free single vortices in each layer,
 whereas the AQ (SQ) phase is a gas of symmetric (anti-symmetric) vortex pairs.
 %, whereas
 %the SQ phase displays free anti-symmetric vortex pairs. 
 %The
 %free vortex pairs make each layer individually appear disordered, 
 %only by considering   

 We now consider the full RG system, including $J_\perp$. 
We numerically integrate the RG equations, and find the phase
diagram 
 %that is
 shown in Fig. \ref{pdint} in terms of the
temperature $T$ and the interaction $J_{\rm int}$. We again
 find four
different phases that are different combinations of LRO,
 QLRO, and disorder in the symmetric and
anti-symmetric sector. 
 At high temperatures we find that both sectors are disordered
in a TBG phase, as before. 
 For lower temperatures, and 
 for a wide range of $J_{{\rm int}}$, 
the system is in a double-layer SF phase (DLSF): 
 The symmetric sector shows algebraic scaling, 
 whereas the exponent of the anti-symmetric sector is renormalized
 to zero, 
 i.e. we find two
SFs that are phase-locked due to 
 %the presence of 
 $J_\perp$. 
%We find
%true SF order in the anti-symmetric sector, and quasi-order in the
%symmetric sector. 
%
 Note that the transition
temperature $T_c$ between DLSF and TBG has been noticeably increased
relative to the decoupled value $T_{KT}$, as we will discuss further
later on. We also find two additional phases, which
 are partially (quasi-)SF and partially disordered. 
 One of them is the SQ phase, as before, whereas the other one (ASF),
 now shows true LRO in the anti-symmetric sector due to 
 %the
 %presence of 
 $J_\perp$, whereas the
symmetric sector remains disordered. 
%The second phase (SQ) is
%characterized by quasi-order in the symmetric sector, but disorder
%in the anti-symmetric sector. 
 We note that the generic double-layer
action that we discuss in this paper does not show a sliding
phase~\cite{toner}, for any non-zero $J_\perp$. Either 
 %the terms
$S_1$ or $S_a$, which is generated by RG, drives the anti-symmetric
sector to a disordered state, or $S_\perp$ creates true LRO in the
field $\phi_a$.

We also use the RG flow to find the order of the phase
transitions in the weak-coupling limit 
 that the anti-symmetric sector undergoes, by determining
the energy gap using a 'poor-man's scaling' argument: when the
coupling amplitude $J_\perp (l^*)$ is of order unity the
corresponding gap is given by the expression $\Delta \sim J_\perp
\exp(-l^*)$. From the behavior of 
 $\Delta$
 %the latter 
 at the phase transition
we can read off whether it is of first or second order, as indicated
in Fig.~\ref{pdint}.

Given the nature 
 %of this framework 
 of an effective theory, only
approximate statements can be made about how the different regimes
of the phase diagram relate to the microscopic interactions. To
create the ASF or AQ phase an attraction between the two
  atom species is needed 
 that is of order $J$, whereas to create the SQ
phase, a repulsion of that order would be needed.
 To detect the different phases, one could use the interference method
 used in \cite{zoran} to distinguish the phase-locked
 phases (DLSF and ASF), which would show a well-defined interference
 pattern, from the uncorrelated phases. Another approach would be
 % to use
 time-of-flight images: The DLSF phase would display a quasi-condensate
 signature, whereas the other phases would appear disordered. However,
 at the transition from ASF 
 %to TBG or from 
 or SQ to TBG, the width of the
 distribution would abruptly increase.

\section{Kibble-Zurek mechanism}\label{kzsection}
In this section we discuss how the phase-locking transition
 found in the previous section could be used to realize the KZ mechanism.
  The defining property of this mechanism 
 is the generation of topological defects
 by ramping across a phase transition, coming from the disordered
 phase. The disordered phase that
 we propose to use 
is the 
  TBG phase
 of the {\it decoupled} 2D systems, that is, we consider
  the experimental setup  reported  in ref.~\cite{zoran} for
 a temperature $T$ above the KT temperature $T_{KT}$.
  The ordered phase we consider is the DLSF phase, i.e. 
  the phase-locked phase of two {\it coupled} SFs. 
 The ramping is achieved by turning on the tunneling between the two layers,
 which can be done by lowering the potential barrier
 between them. 
  For this procedure the critical 
 temperature $T_c$ of the DLSF-TBG transition needs to be
 above the KT temperature of the uncoupled systems.
  We now show that the RG flow indeed predicts such a scenario. 
 In the experiments in Ref.~\cite{zoran}, the atoms
in different layers do not interact with each other. Therefore, it
can be expected that $J_{\rm int}$ is small,  of order 
 $J_\perp$, 
 which motivates
 us to discuss the case $J_{int}=0$ here. 
 We note however, that the desired scenario of an increased
 critical temperature,  
 %of the coupled system compared to the uncoupled
 %system, 
 is found for a wide range of $J_{int}$, as
 can be seen in Fig. \ref{pdint}. 
 In Fig. \ref{PD} (a) we show how the critical temperature
of the DLSF-TBG transition behaves, predicted by the RG flow, for
different values of $A_1$.
 %, for $J_{\rm int}=0$. 
 The critical temperature shows a sizeable
increase, due to the phase-locking transition. Due to the
perturbative nature of the RG scheme, the RG flow underestimates the
effects of the term $S_1$, and predicts a finite jump of the
critical temperature when $J_\perp$ is turned on. However, to lock
the SFs together in the regime slightly above $T_{KT}$,
$J_\perp$ needs to be at least of the order of the vortex core
energy, giving rise to a finite slope of $T_c$ instead of a jump.
The energy gap of this transition is shown in Fig.~\ref{PD} from
which we can see that the transition is of first order, 
 in contrast to the second order transition 
 described in \cite{kibble, zurek}, which
 is advantageous because the onset of order is instantaneous
 rather than continuous. 
 We
note that the phase diagram was obtained using the assumption that
the bare parameters of the model, in particular 
%the stiffness
%parameter 
 $J$, do not depend on temperature. This is true only if
temperature is close to $T_{KT}$. Here we find that the ratio
$T_c/T_{KT}$ can be relatively large. In fact $T_c/T_{KT}$ will be
always smaller than that shown in Fig~\ref{PD} (a), however,
qualitatively the behavior of $T_c/T_{KT}$ as a function of
$J_{\perp}$ should remain intact. We point out that our results can be
generalized to a system of $N>2$ coupled SFs. One finds that the SFs
still show a strong tendency to phase-lock together. As a result
the critical temperature should approximately satisfy the equality
$\pi J(T_c) N =2 T_c$. Thus as $N$ increases $T_c$ approaches the
mean-field critical temperature at which the stiffness $J$ vanishes
and we recover the usual 3D result.

In finite size systems there is another constraint on the minimum
value of $J_{\perp}$:
% required to develop a jump in $T_c$. 
 %Indeed,
 We consider the free energy of a single vortex in the anti-symmetric
field: $\phi_a \sim \arctan(x/y)$. For the decoupled system we get
 for the free energy \cite{SG}: 
 $F \sim 2(\pi J -2 T)\log L/\alpha$, where
 $L$ is the system size. 
 The coupling term
gives a free energy 
 contribution $F_{\perp}\sim J_\perp (L/\alpha)^2$. In the
thermodynamic limit, $L \rightarrow \infty$, this term diverges
faster than the others, which is consistent with our finding of LRO
in the antisymmetric sector. For a finite system,  comparing 
these terms gives the estimate $J_\perp\sim J
\log(L/\alpha)/(L/\alpha)^2$, that is required for this order to
develop.
 With a system size $L/\alpha \gtrsim 10^2$, that 
 would require $J_\perp\gtrsim 10^{-3} J$, which, for the 
 setup in \cite{zoran}, would be around $10^2 s^{-1}$.

As an estimate of the number  of domains that would be created, we
 follow the argument in \cite{kibble}:
 The coherence scale of the DLSF phase is given by $(\Delta/J)^{1/2}\alpha$,
 which is the scale 
 %that this defined by 
  of a Klein-Gordon model with
 a kinetic energy scale $J$ and a 'mass-term' with a prefactor
 $\Delta/\alpha^2$. 
 The domain size is then given by $(\Delta/J)\alpha^2$, and the
 number of domains 
 %in the system 
 by $\sim (J/\Delta) L^2/\alpha^2$.
  As we show in Fig. \ref{PD} b) for $J_\perp/J\approx 10^{-2}$, 
 we find $\Delta/J_\perp \sim 10^{-1}$, and therefore 
 $J/\Delta \sim 10^3$. With $L/\alpha \gtrsim 10^2$,
  we would get $N_{dom} \sim 10^1 - 10^2$, which would generate a 
 similar number of vortices. 
  We estimate the vortex-antivortex imbalance by considering
 the number of domains around the periphery of the 
 system, which scales as $L/\xi$. If we imagine that the
 phase behaves like a random walk, the total phase mismatch, corresponding
 to the vortex-antivortex imbalance, 
 will scale as $\sqrt{L/\xi} \sim N_{dom}^{1/4}$, 
 which, for $L/\alpha\gtrsim 10^2$,
 is of the order $10^0 - 10^1$.

In summary, we propose the following procedure: 
 i) Prepare two uncoupled SFs at a temperature $T$ slightly above 
 $T_{KT}$.
  ii) Switch on the tunneling between the two layers, which creates a DLSF 
 phase with a critical temperature $T_c$ higher than $T$.
 As a result, one should find a number of long-lived 
 vortex-antivortex pairs in the 
 anti-symmetric phase field $\phi_a$, which would 
 be visible in an interference measurement, at
 a temperature where there would be none in thermal equilibrium. 
% \revision{The defects in the field $\phi_a$ would be visible }

%As was demonstrated in Ref.~[\cite{zoran}], the KT transition
%is an unbinding transition of vortex-anti-vortex pairs. This
%transition is therefore indicated and driven by the appearance of
%free vortices in the system that destroy the QLRO. In
%ref.~[\cite{zoran}] this transition could be seen directly in
%the time of flight (TOF) images: in the interference picture of two
%neighboring 2D systems, where one of the two contains a vortex and
%the other one does not, a phase jump of $\pi$ is visible, which
%indicates the presence of a vortex. For the phase-locking transition
%discussed in this paper, such a signature would be absent in the
%phase-locked state, because in that state the phase difference
%between the two SFs is fixed at a multiple of $2\pi$. The phase
%transition that occurs for this DLSF, at the renormalized transition
%temperature, is a KT transition in the charge sector. Therefore, the
%vortices that drive this transition are vortices in the total
%density, and are therefore visible in a TOF image taken along the
%perpendicular direction of the system.

\begin{figure}[t]
\onefigure[width=4.3cm]{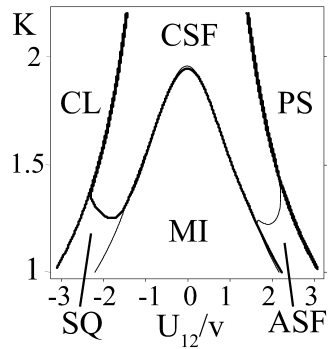}
\caption{\label{pdint1} Phase diagram of two coupled lattice 1D
bosonic SFs at unit filling: Luttinger parameter $K$ versus
interaction (in units of SF velocity). MI: Mott insulator;
SQ: symmetric quasi-order; PS: phase separation; CSF:
coupled superfluids; CL: collapsed phase; ASF: anti-symmetric
 SF.}
\end{figure}

\section{1D superfluids}
%
%
%
%Next 
 %In this section 
 We now study 
%we consider 
 the phases of 
 two coupled SFs in 1D (see \cite{donohue, ho, vekua}). 
%periodic potential,
 We  consider 
%systems 
%  with either commensurate filling, i.e.
  both a 
 lattice system with unit filling, and a system
 with  incommensurate filling.
 %, i.e.
 %either continuous systems or lattice systems
 %with filling away from unit filling.
 These types of
systems are formally  closely related to the 2D system that we
discussed before~\cite{giamarchi_book}. The imaginary time direction
in a quantum 1D system plays the role of the second dimension in the
2D case. 

The action of two incommensurate SFs, coupled
 by a hopping term, is given by 
%
%
%%
%\bea\label{Sinc}
 $S  =  S_{0,1} + S_{0,2} + S_{\perp}$.
%\eea
%
In the terms $S_{0,1/2}$ we formally replace the ratio $\pi J/T$ 
 by the Luttinger parameter $K$, 
 and similarly we introduce $g_\perp=J_\perp/T$ in the
 hopping term.
%\bea
%S_{0, j} & = & \frac{1}{2\pi K} \int d^2r [(\partial_{\tau} \theta_{j})^2
%+   (\partial_x \theta_{j})^2] \, .
%\eea
%
%These SFs are characterized by a Luttinger 
%
 The 
parameter $K$ and a
velocity $v$, which is contained in  ${\bf r}=(v \tau,
x)$, characterize
 these 1D SFs, where $K$ 
% The Luttinger parameter
  is a measure of the interaction:
 weakly interacting bosons have $K\gg 1$,
 while bosons interacting via
 a strongly repulsive potential, have $K\to 1$ (Ref.~\cite{cazalilla}). 
%The hopping term between these two
%systems is given by:
%\bea\label{hop_1D}
%S_\perp & = &  2 g_\perp \int \frac{d^2 r}{(2 \pi\alpha)^2}
%\cos(\phi_1-\phi_2) \, .
%\eea
The system again separates into symmetric and anti-symmetric fields,
Eq.~(\ref{spincharge}), where the anti-symmetric
 sector is given by a sine-Gordon model.
 By going to this basis, one can determine
that the scaling dimension of the operator $\cos(\sqrt{2}\phi_a)$ is
given by $1/(2 K)$. Therefore for any value of $K$, this term is
highly relevant, and the anti-symmetric sector develops LRO, 
 $K_a\rightarrow \infty$, giving rise to a phase of coupled
SFs (CSF). 

For 
 %a system of commensurate SFs, i.e. 
 a lattice
 system of SFs 
 %where each SF has 
 with unit filling, this
 phase-locking tendency competes with the Mott insulator transition,
 and
 %, as we will see,
  other, partial, localization transitions.
 We include  
 the terms $S_1$ and $S_{s,a}$
%contributions 
 in the
action, 
 where we replace $A_1/T$ by $g_1$, and $A_{s,a}/T$ by
 $g_{s,a}$,
 and finally we add 
 %the term 
%
%\bea\label{S1_1D} S_1 & = &  2 g_1 \int \frac{d^2 r}{(2
%\pi\alpha)^2} [\cos(2\theta_1) + \cos(2\theta_2)] \, ,
%\eea
%and again the terms of the next order:
%\bea\label{Srs_1D}
%S_{s/a} & = &  2 g_{s/a} \int \frac{d^2 r}{(2 \pi\alpha)^2} \cos(2\sqrt{2}\theta_{s/a}) \, .
%\eea
%
%
%
%
% As in the 2D case, we include an interaction term of the
% form:
%If the bosons have some  then there is an additional term in the
%action:
%\bea
$S'_{12}  =  (U_{12}/\pi^2) \int d^2 r \nabla \theta_1 \nabla
\theta_2$
%\eea
%
 which would derive from some inter-chain
density-density interaction. 
%
%However this term alone does not qualitatively affect the flow of
%$g_\perp$ to large values and thus the system remains in the CSF
%phase.
%
From these expressions it is clear that the 1D commensurate system
is formally identical to the 2D system at finite temperature. 
% The RG equations can be obtained 
% by matching the coefficients
% of the 2D action to the LL action. 
% From 
 We again integrate the RG equations
numerically,  
%integrating
% these equations, 
  and we find that
%We
%thus can immediately deduce the phase diagram from the previous
%analysis. Thus 
 if the parameters $g_\perp$ and $g_1$ are of
comparable magnitude then the hopping term is the most relevant
one, driving the system to 
 a CSF phase. 
%the limit of coupled SFs. 
 However, for $g_1
\gg g_\perp$, and for
 % strongly interacting bosons, i.e.
  small $K$,
  the terms $S_1, S_s$, or $S_a$ can dominate, and
localize either the symmetric or the anti-symmetric sector, or both.
The resulting phase diagram is very similar to that of the 2D system
with the localized phases corresponding to the thermal ones (see
Fig.~\ref{pdint1}).

%The RG flow equations are  now given by:
%\bea\label{RG_1D}
%\frac{d g_\perp}{d l} & = &(2-\frac{1}{2K_a}) g_\perp \, ,
%\nonumber
%\\
%\frac{d g_s}{d l} & = &(2-2K_s) g_s + \alpha_3\frac{g_1^2 (K_a -K_s)}{2 \pi}
%\, ,
%\nonumber
%\\
%\frac{d g_a}{d l} & = &(2-2K_a) g_a + \alpha_3\frac{g_1^2 (K_s -K_a)}{2 \pi}
%\, ,
%\nonumber
%\\
%\frac{d g_1}{d l} & = &(2-\frac{K_s+K_a}{2} + \alpha_3\frac{g_s K_s + g_a
%  K_a}{\pi}) g_1 \, ,
%\nonumber
%\\
%\frac{d K_a}{d l} & = & \alpha_2 \Big(\frac{g_\perp^2}{4\pi^2}\frac{1}{K_a} -
%4\frac{g_{a}^2}{\pi^2}  K_a^3 - \frac{g_1^2}{2 \pi^2} (K_s+K_a)K_a^2\Big) \, ,
%\nonumber
%\\
%\frac{d K_s}{d l} & = & - \alpha_2 \Big(2\frac{g_{s}^2}{\pi^2}  K_s^2 +
%\frac{g_1^2}{4\pi^2} (K_s+K_a) K_s \Big)2 K_s \, .
%\nonumber
%\eea

%For consistency, we expand the parameters $K_{s/a}$ as:
%\bea
%K_{s/a} & = & (1/K \pm U_{12}/\pi)^{-1} + \epsilon_{s/a} \, .
%\eea

\section{Conclusion}
In summary, we studied the phase-locking transition of 1D and 2D
 SFs, within an RG approach. For 2D SFs,
this transition is accompanied by an increase of the transition
temperature. We suggest that this effect can be used to probe the
 KZ mechanism in cold atom systems by rapidly changing the
ratio $T_c/T$. For both 2D and commensurate 1D systems, interaction
terms give rise to additional phases, in which  either the symmetric
or the anti-symmetric sector is disordered or localized,
respectively.

% \section{Section title}
% Insert here the text.
% See fig.~\ref{fig.1}, table~\ref{tab.1} and eq.~(\ref{eq.1}).
% See also~\cite{b.a,b.b}.
% \begin{equation}
% \label{eq.1}
% 0\neq1
% \end{equation}

%\begin{figure}
%\onefigure{epl-template.eps}
%\caption{Figure caption.}
%\label{fig.1}
%\end{figure}

% \begin{table}
% \caption{Table caption.}
% \label{tab.1}
% \begin{center}
% \begin{tabular}{lcr}
% first  & table & row\\
% second & table & row
% \end{tabular}
% \end{center}
% \end{table}

\acknowledgments
We acknowledge illuminating discussions with B. Halperin, S.
Sachdev, and Z. Hadzibabic. A.~H.~C.~N. was supported by the NSF
grant DMR-0343790, A.P. was supported by AFOSR YIP.


\begin{thebibliography}{0}
\bibitem{kibbleremark}
% In the meantime 
 Since then, it has been established that cosmic strings can only 
 contribute a small fraction of the initial density perturbations. We
 thank T. Kibble for this comment.
\end{thebibliography}
\end{document}